\documentclass{tlp}
\usepackage{epsfig}
\usepackage[utf8]{inputenc}
\usepackage{multicol}
\usepackage{multirow}
\usepackage{wrapfig}
\newcommand{\verbatimproperties}{\renewcommand{\baselinestretch}{0.85} \small}

\begin{document}

\label{firstpage}

\title[On Combining Linear-Based Strategies for Tabled Evaluation]
      {On Combining Linear-Based Strategies for Tabled Evaluation of Logic Programs}

\author[Miguel Areias and Ricardo Rocha]
       {MIGUEL AREIAS and RICARDO ROCHA\\
       CRACS \& INESC-Porto LA, Faculty of Sciences, University of Porto\\
       Rua do Campo Alegre, 1021/1055, 4169-007 Porto, Portugal\\
       \email{\{miguel-areias,ricroc\}@dcc.fc.up.pt}}

\maketitle


\begin{abstract}
  Tabled evaluation is a recognized and powerful technique that
  overcomes some limitations of traditional Prolog systems in dealing
  with recursion and redundant sub-computations. We can distinguish
  two main categories of tabling mechanisms: suspension-based tabling
  and linear tabling. While suspension-based mechanisms are considered
  to obtain better results in general, they have more memory space
  requirements and are more complex and harder to implement than linear
  tabling mechanisms. Arguably, the SLDT and DRA strategies are the
  two most successful extensions to standard linear tabled
  evaluation. In this work, we propose a new strategy, named DRS, and
  we present a framework, on top of the Yap system, that supports the
  combination of all these three strategies. Our implementation shares
  the underlying execution environment and most of the data structures
  used to implement tabling in Yap. We thus argue that all these
  common features allows us to make a first and fair comparison
  between these different linear tabling strategies and, therefore,
  better understand the advantages and weaknesses of each, when used
  solely or combined with the others.
\end{abstract}



\section{Introduction}

The operational semantics of Prolog is given by SLD
resolution~\cite{Lloyd-87}, an evaluation strategy particularly simple
that matches current stack based machines particularly well, but that
suffers from fundamental limitations, such as in dealing with
recursion and redundant sub-computations. Tabled
evaluation~\cite{Tamaki-86,Chen-96} is a recognized and powerful
technique that can considerably reduce the search space, avoid looping
and have better termination properties than SLD resolution.

Tabling consists of storing intermediate solutions for subgoals so
that they can be reused when a repeated subgoal appears during the
resolution process. Implementations of tabling are currently available
in systems like XSB Prolog~\cite{Sagonas-98}, Yap
Prolog~\cite{Rocha-00a}, B-Prolog~\cite{Zhou-00},
ALS-Prolog~\cite{Guo-01}, Mercury~\cite{Somogyi-06} and Ciao
Prolog~\cite{Guzman-09a}. In these implementations, we can distinguish
two main categories of tabling mechanisms: \emph{suspension-based
  tabling} and \emph{linear tabling}.

Suspension-based tabling mechanisms need to preserve the computation
state of suspended tabled subgoals in order to ensure that all
solutions are correctly computed. A tabled evaluation can be seen as a
sequence of sub-computations that suspend and later resume. Linear
tabling mechanisms use iterative computations of tabled subgoals to
compute fix-points and for that they maintain a single execution tree
without requiring suspension and resumption of sub-computations. While
suspension-based mechanisms are considered to obtain better results in
general, they have more memory space requirements and are more complex
and harder to implement than linear tabling mechanisms.

Arguably, the SLDT~\cite{Zhou-00} and DRA~\cite{Guo-01,Areias-10}
strategies are the two most successful extensions to standard linear
tabling evaluation. As these strategies optimize different aspects of
the evaluation, they are, in principle, orthogonal to each other and
thus it should be possible to combine both in the same
system. However, to the best of our knowledge, no single Prolog system
supports both strategies simultaneously and thus, understanding the
advantages and weaknesses of each cannot be fully dissociated from the
base Prolog system on top of which they are implemented.

In this work, we propose a new strategy, named \emph{Dynamic
  Reordering of Solutions (DRS)}, and we present a framework, on top
of the Yap Prolog system, that integrates and supports the combination
of the SLDT, DRA and DRS strategies. Our implementation shares the
underlying execution environment and most of the data structures used
to implement tabling in Yap~\cite{Rocha-00a}. In particular, we took
advantage of Yap's efficient table space data structures based on
\emph{tries}~\cite{RamakrishnanIV-99}, which we used with minimal
modifications. We thus argue that all these common support features
allows us to make a first and fair comparison between these different
linear tabling strategies and, therefore, better understand the
advantages and weaknesses of each, when used solely or combined with
the others.

The remainder of the paper is organized as follows. First, we briefly
introduce the basics of tabling and describe the execution model for
standard linear tabled evaluation. Next, we present the SLDT, DRA and
DRS strategies and discuss how they can be used to optimize different
aspects of the evaluation. We then provide some implementation details
regarding the integration of the three strategies on top of the Yap
engine. Finally, we present some experimental results and we end by
outlining some conclusions.


\section{Standard Linear Tabled Evaluation}

Tabling works by storing intermediate solutions for tabled subgoals so
that they can be reused when a repeated call appears\footnote{A
  subgoal call repeats a previous call if they are the same up to
  variable renaming.}. In a nutshell, first calls to tabled subgoals
are considered \emph{generators} and are evaluated as usual, using SLD
resolution, but their solutions are stored in a global data space,
called the \emph{table space}. Repeated calls to tabled subgoals are
considered \emph{consumers} and are not re-evaluated against the
program clauses because they can potentially lead to infinite loops,
instead they are resolved by consuming the solutions already stored
for the corresponding generator. During this process, as further new
solutions are found, we need to ensure that they will be consumed by
all the consumers, as otherwise we may miss parts of the computation
and not fully explore the search space.

A generator call $C$ thus keeps trying its matching clauses until a
fix-point is reached. If no new solutions are found during one cycle
of trying the matching clauses, then we have reached a fix-point and
we can say that $C$ is completely evaluated. However, if a number of
subgoal calls is mutually dependent, thus forming a \emph{Strongly
  Connected Component (SCC)}, then completion is more complex and we
can only complete the calls in a SCC together~\cite{Sagonas-98}. SCCs
are usually represented by the \emph{leader call}, i.e., the generator
call which does not depend on older generators. A leader call defines
the next completion point, i.e., if no new solutions are found during
one cycle of trying the matching clauses for the leader call, then we
have reached a fix-point and we can say that all subgoal calls in the
SCC are completely evaluated.

\begin{figure}[ht]
\centering
\includegraphics[width=12cm]{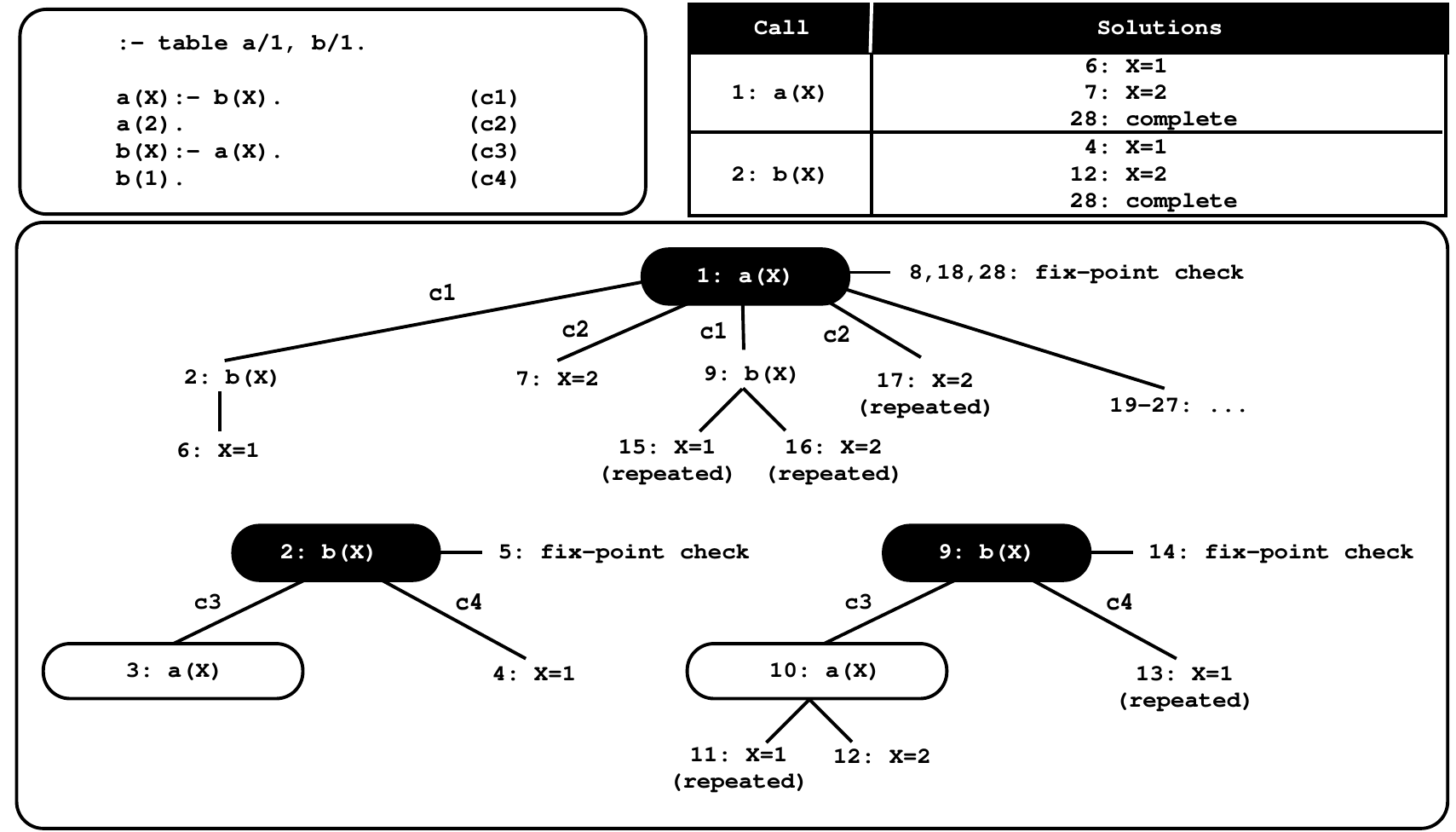}
\caption{A standard linear tabled evaluation}
\label{fig_linear_evaluation}
\end{figure}

We next illustrate in Fig.~\ref{fig_linear_evaluation} the standard
execution model for linear tabling. At the top, the figure shows the
program code (the left box) and the final state of the table space
(the right box). The program defines two tabled predicates,
\texttt{a/1} and \texttt{b/1}, each defined by two clauses (clauses
\texttt{c1} to \texttt{c4}). The bottom sub-figure shows the
evaluation sequence for the query goal \texttt{a(X)}. Generator calls
are depicted by black oval boxes and consumer calls by white oval
boxes.

The evaluation starts by inserting a new entry in the table space
representing the generator call \texttt{a(X)} (step 1). Then,
\texttt{a(X)} is resolved against its first matching clause, clause
\texttt{c1}, calling \texttt{b(X)} in the continuation. As this is a
first call to \texttt{b(X)}, we insert a new entry in the table space
representing \texttt{b(X)} and proceed as shown in the bottom left
tree (step 2). Subgoal \texttt{b(X)} is also resolved against its
first matching clause, clause \texttt{c3}, calling again \texttt{a(X)}
in the continuation (step 3). Since \texttt{a(X)} is a repeated call,
we try to consume solutions from the table space, but at this stage no
solutions are available, so execution fails.

We then try the second matching clause for \texttt{b(X)}, clause
\texttt{c4}, and a first solution for \texttt{b(X)}, \texttt{X=1}, is
found and added to the table space (step 4). We then follow a
\emph{local scheduling} strategy and execution
\emph{fails}~\cite{Freire-96}. With local scheduling, new solutions
are only returned to the calling environment when all program clauses
were explored. The execution thus fails back to node 2 and we check
for a fix-point (step 5), but \texttt{b(X)} is not a leader call
because it has a dependency (consumer node 3) to an older call,
\texttt{a(X)}. Remember that we reach a fix-point when no new
solutions are found during the last cycle of trying the matching
clauses for the leader call.

Next, as we are following a local scheduling strategy, the solution
for \texttt{b(X)} should now be propagated to the context of the
previous call. We thus propagate the solution \texttt{X=1} to the
context of the generator call for \texttt{a(X)}, which originates a
first solution for \texttt{a(X)}, \texttt{X=1} (step 6). Then, we try
the second matching clause for \texttt{a(X)} and a second solution,
\texttt{X=2}, is found and added to the table space (step 7). We then
backtrack again to the generator call for \texttt{a(X)} and because we
have already explored all matching clauses, we check for a fix-point
(step 8). We have found new solutions for both \texttt{a(X)} and
\texttt{b(X)}, thus the current SCC is scheduled for re-evaluation.

The evaluation then repeats the same sequence as in steps 2 and 3 (now
steps 9 and 10), but at this time the consumer call for \texttt{a(X)}
has solutions in the table. Solution \texttt{X=1} is first forwarded to
it, which originates a repeated solution for \texttt{b(X)} (step 11)
and thus execution fails. Then, solution \texttt{X=2} is also forward
to it and a new solution for \texttt{b(X)} is found. In the
continuation, we find another repeated solution for \texttt{b(X)}
(step 13) and we fail a second time in the fix-point check for
\texttt{b(X)} (step 14). Again, as we are following a local scheduling
strategy, the solutions for \texttt{b(X)} are propagated to the
context of the generator call for \texttt{a(X)}, but only repeated
solutions are found (steps 15 and 16). Clause \texttt{c2} is then
explored, but without any further developments (step 17).

We then backtrack one more time to the generator call for
\texttt{a(X)} and because we have found a new solution for
\texttt{b(X)} during the last iteration, the current SCC is
scheduled again for re-evaluation (step 18). The re-evaluation of the
SCC does not find new solutions for both \texttt{a(X)} and
\texttt{b(X)} (steps 19 to 27). Thus, when backtracking again to
\texttt{a(X)} we have reached a fix-point and because \texttt{a(X)} is
a leader call, we can declare the two subgoal calls to be completed
(step 28).


\section{Linear Tabling Strategies}

The standard linear tabling mechanism uses a naive approach to
evaluate tabled logic programs. Every time a new solution is found
during the last round of evaluation, the complete search space for the
current SCC is scheduled for re-evaluation. However, some branches of
the SCC can be avoided, since it is possible to know beforehand that
they will only lead to repeated computations, hence not finding any
new solutions. Next, we will present three different strategies for
optimizing the standard linear tabled evaluation. The common goal of
these strategies is to minimize the number of branches to be explored,
thus reducing the search space, and each strategy tries to focus on
different aspects of the evaluation to achieve it.


\subsection{Dynamic Reordering of Execution}

\begin{figure}[ht]
\centering
\includegraphics[width=12cm]{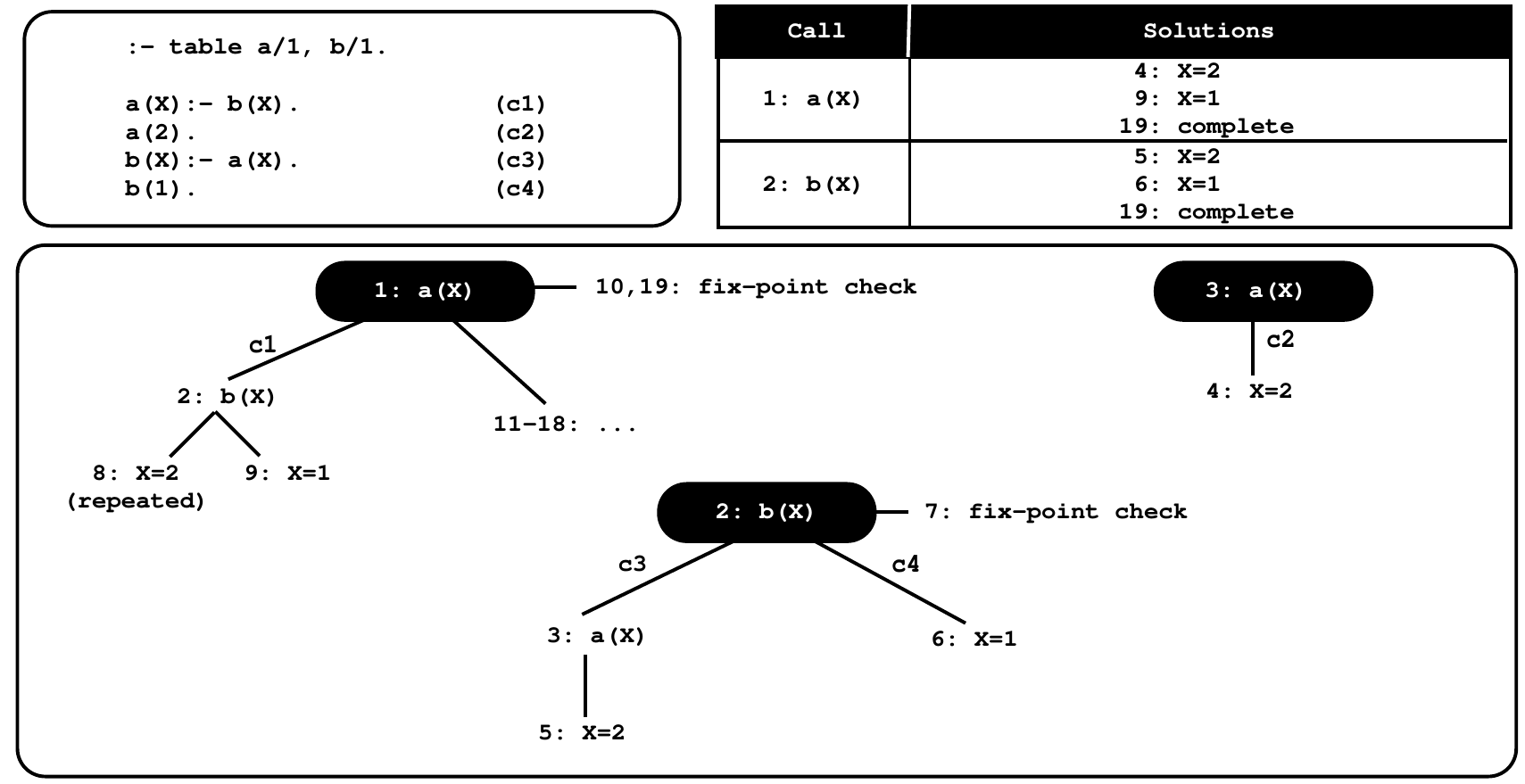}
\caption{Using the DRE strategy to evaluate the program in
  Fig.~\ref{fig_linear_evaluation}}
\label{fig_dre_evaluation}
\end{figure}

The first optimization, that we call \emph{Dynamic Reordering of
  Execution (DRE)}, is based on the original SLDT strategy, as
proposed by Zhou et al.~\cite{Zhou-00}. The key idea of the DRE
strategy is to give priority to the program clauses instead of
consuming answers, and to achieve that it lets repeated calls to
tabled subgoals execute from the \emph{backtracking clause of the
  former call}. A first call to a tabled subgoal is called a
\emph{pioneer} and repeated calls are called \emph{followers} of the
pioneer. When backtracking to a pioneer or a follower, we use the same
strategy, first we explore the remaining clauses and only then we try
to consume solutions. The fix-point check operation is still only
performed by pioneer calls. Figure~\ref{fig_dre_evaluation} uses the
same example from Fig.~\ref{fig_linear_evaluation} to illustrate how
DRE evaluation works.

As for the standard strategy, the evaluation starts with first
(pioneer) calls to \texttt{a(X)} (step 1) and \texttt{b(X)} (step 2),
and then, in the continuation, \texttt{a(X)} is called repeatedly
(step 3). But now with DRE evaluation, \texttt{a(X)} is considered a
follower and thus we \emph{steal} the backtracking clause of the
former call at node 1, i.e., the second matching clause for
\texttt{a(X)}, clause \texttt{c2}. The evaluation then proceeds as for
a generator call (right upper tree in Fig.~\ref{fig_dre_evaluation}),
which means that new solutions can be generated for \texttt{a(X)}. We
thus try \texttt{c2}, and a first solution for \texttt{a(X)},
\texttt{X=2}, is found and added to the table space (step 4).  We then
follow a local scheduling strategy and execution fails backtracking to
the follower node. As both matching clauses for \texttt{a(X)} were
already taken, we try to consume solutions. The solution \texttt{X=2}
is then propagated to the context of \texttt{b(X)} which originates
the solution \texttt{X=2} (step 5). Next, in step 6 we find the second
solution for \texttt{b(X)} and in step 7 we check for a fix-point, but
\texttt{b(X)} is not a leader call because it has a dependency
(follower node 3) to an older call, \texttt{a(X)}. The solutions for
\texttt{b(X)} are then propagated to the context of the pioneer call
for \texttt{a(X)}, which originates a second solution for
\texttt{a(X)}, \texttt{X=1} (step 9). We then backtrack to the pioneer
call for \texttt{a(X)} and because we have already explored the
matching clause \texttt{c2} in the follower node 3, we check for a
fix-point. Since we have found new solutions during the last
iteration, the current SCC is scheduled for re-evaluation (step
10). The re-evaluation of the SCC does not find any further solutions
(steps 11 to 18), and thus the evaluation can be completed at step 19.


\subsection{Dynamic Reordering of Alternatives}

The key idea of the \emph{Dynamic Reordering of Alternatives (DRA)}
strategy, as originally proposed by Guo and Gupta~\cite{Guo-01}, is to
memoize the clauses (or alternatives) leading to consumer calls, the
\emph{looping alternatives}, in such a way that when scheduling an SCC
for re-evaluation, instead of trying the full set of matching clauses,
we only try the looping alternatives.

\begin{figure}[ht]
\centering
\includegraphics[width=12cm]{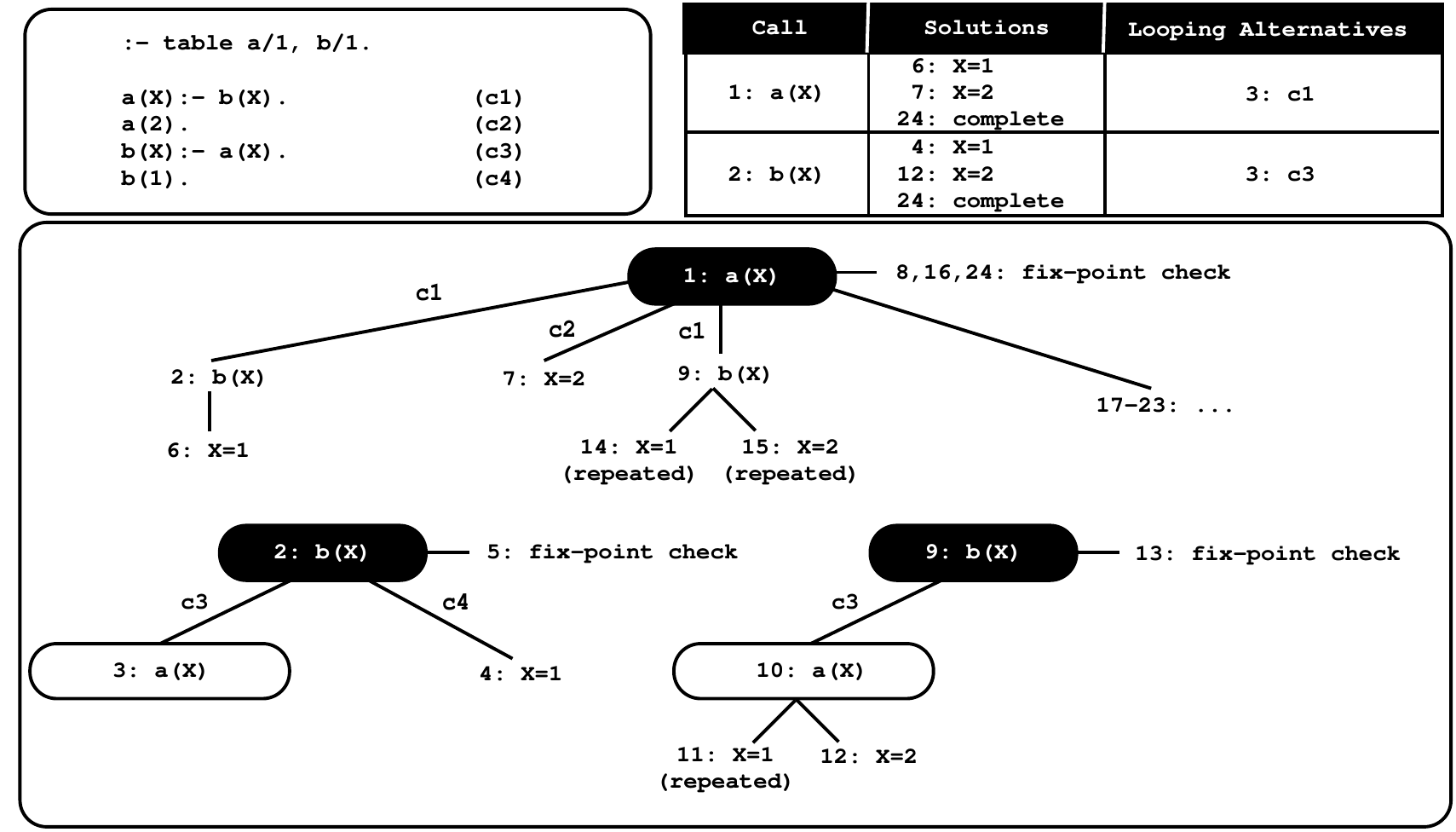}
\caption{Using the DRA strategy to evaluate the program in
  Fig.~\ref{fig_linear_evaluation}}
\label{fig_dra_evaluation}
\end{figure}

Initially, a generator call $C$ explores the matching clauses as in
standard linear tabled evaluation and, if a consumer call is found,
the current clause for $C$ is memoized as a looping alternative. After
exploring all the matching clauses, $C$ enters the \emph{looping
  state} and from this point on, it only tries the looping
alternatives until a fix-point is
reached. Figure~\ref{fig_dra_evaluation} uses again the same example
from Fig.~\ref{fig_linear_evaluation} to illustrate how DRA evaluation
works.

The evaluation sequence for the first SCC round (steps 2 to 7) is
identical to the standard evaluation of
Fig.~\ref{fig_linear_evaluation}. The difference is that this round is
also used to detect the alternatives leading to consumers calls. We
only have one consumer call at node 3 for \texttt{a(X)}. The clauses
in evaluation up to the corresponding generator, call \texttt{a(X)} at
node 1, are thus marked as looping alternatives and added to the
respective table entries. This includes alternative \texttt{c3} for
\texttt{b(X)} and alternative \texttt{c1} for \texttt{a(X)}.  As for
the standard strategy, the SCC is then scheduled for two extra
re-evaluation rounds (steps 9 to 15 and steps 17 to 23), but now only
the looping alternatives are evaluated, which means that the clauses
\texttt{c2} and \texttt{c4} are ignored.


\subsection{Dynamic Reordering of Solutions}

\begin{figure}[ht]
\centering
\includegraphics[width=12cm]{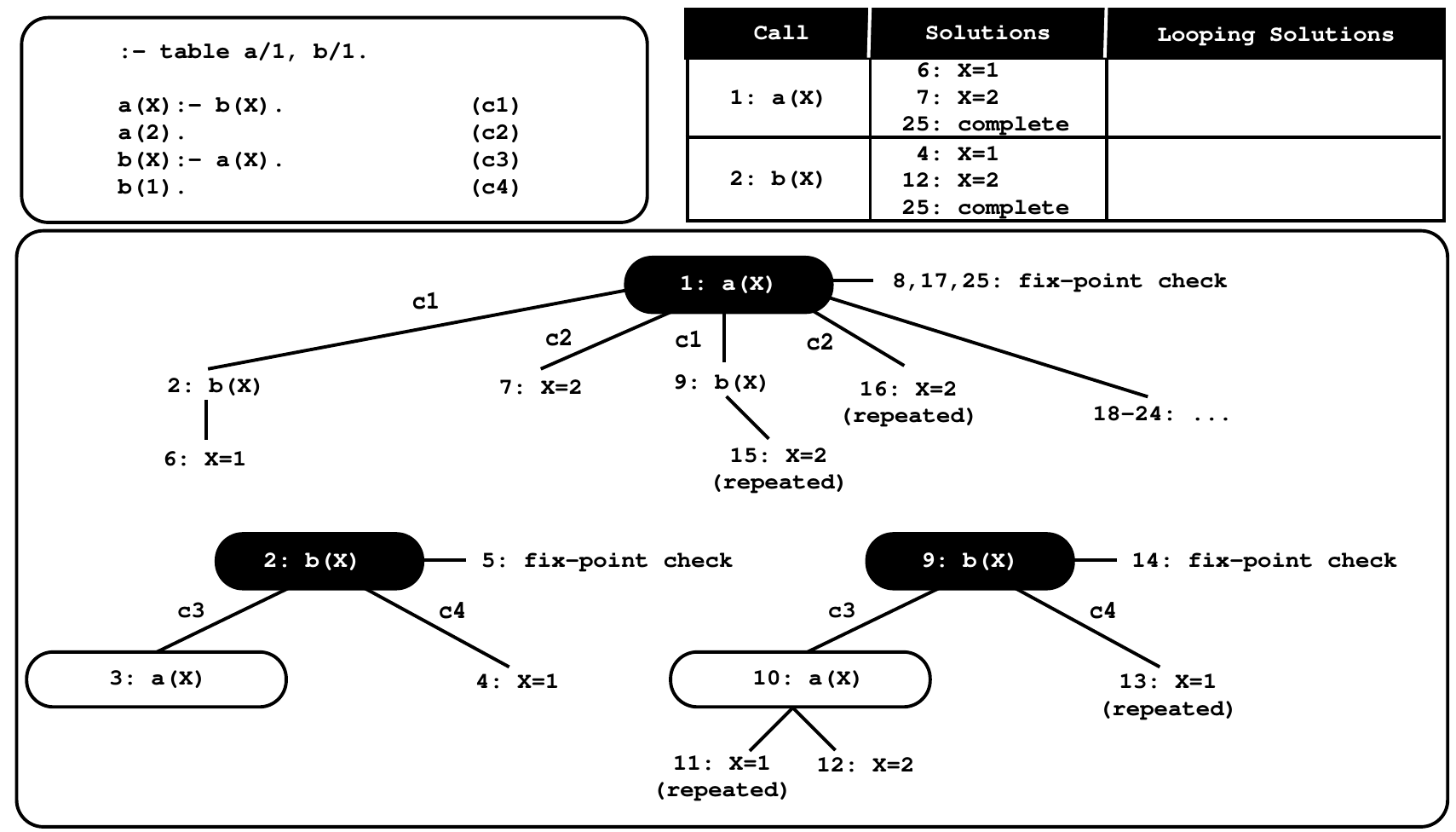}
\caption{Using the DRS strategy to evaluate the program in
  Fig.~\ref{fig_linear_evaluation}}
\label{fig_drs_evaluation}
\end{figure}

The last optimization, that we named \emph{Dynamic Reordering of
  Solutions (DRS)}, is a new proposal that can be seen as a variant of
the DRA strategy, but applied to the consumption of solutions. The key
idea of the DRS strategy is to memoize the solutions leading to
consumer calls, the \emph{looping solutions}. When a non-leader
generator call $C$ consumes solutions to propagate them to the context
of the previous call, if a consumer call is found, the current
solution for $C$ is memoized as a looping solution. Later, if $C$ is
scheduled for re-evaluation, instead of trying the full set of
solutions, it only tries the looping solutions plus the new solutions
found during the current round. In each round, the new solutions
leading to consumer calls are added to the previous set of looping
solutions. In Fig.~\ref{fig_drs_evaluation}, we use again the same
example from Fig.~\ref{fig_linear_evaluation} to illustrate how DRS
evaluation works.

In this example, we only have one non-leader generator call,
\texttt{b(X)}, which is called once for each evaluation round over the
SCC (steps 2, 9 and 18 in Fig.~\ref{fig_drs_evaluation}). By following
the evaluation, it is possible to verify that no solutions are marked
as looping solutions, and thus, on each round, \texttt{b(X)} only
consumes the new solutions found during the round. This means that
solution \texttt{X=1} only is consumed on the first round (step 6),
solution \texttt{X=2} only is consumed on the second round (step 15)
and no solution is consumed on the last round.


\section{Implementation Details}

This section describes some implementation details regarding the
integration of the three strategies on top of the Yap engine, with
particular focus on the table space data structures and on the tabling
operations.


\subsection{Table Space}

\begin{wrapfigure}{!r}{9cm}
\centering
\includegraphics[width=9cm]{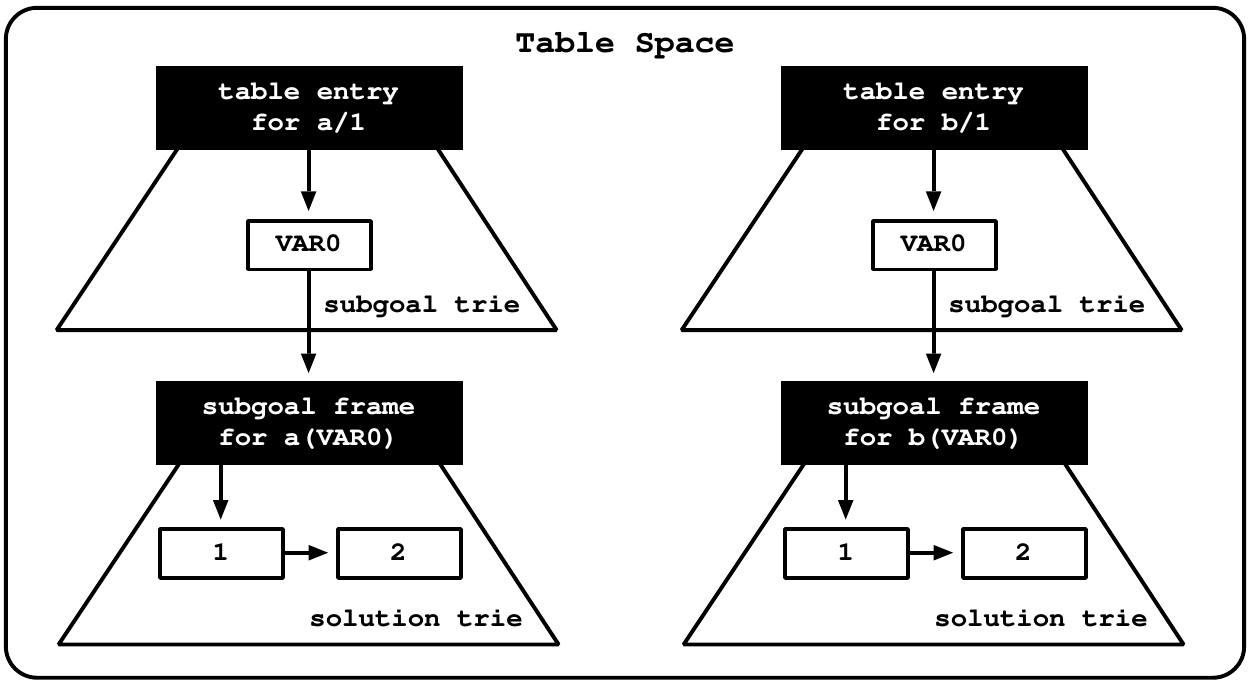}
\caption{Table space organization}
\label{fig_table_space}
\end{wrapfigure}

To implement the table space, Yap uses \emph{tries} which is
considered a very efficient way to implement the table
space~\cite{RamakrishnanIV-99}. Tries are trees in which common
prefixes are represented only once. Tries provide complete
discrimination for terms and permit look up and insertion to be done in
a single pass. Figure~\ref{fig_table_space} details the table space
organization for the example used on the previous sections.

As other tabling engines, Yap uses two levels of tries: one for the
subgoal calls and other for the computed solutions. A tabled predicate
accesses the table space through a specific \emph{table entry} data
structure. Each different subgoal call is represented as a unique path
in the \emph{subgoal trie} and each different solution is represented
as a unique path in the \emph{solution trie}. A key data structure in
this organization is the \emph{subgoal frame}. Subgoal frames are used
to store information about each tabled subgoal call, namely: the entry
point to the solution trie; the state of the subgoal (\emph{ready},
\emph{evaluating} or \emph{complete}); support to detect if the
subgoal is a leader call; and support to detect if new solutions were
found during the last round of evaluation. The DRE, DRA and DRS
strategies extend the subgoal frame data structure with the following
extra information:

\begin{description}
\item [DRE:] the pioneer call; and the backtracking clause of the
  former call.
\item [DRA:] support to detect, store and load looping alternatives;
  and two new states, \emph{loop\_ready} and \emph{loop\_evaluating},
  used to detect, respectively, generator and consumer calls in
  re-evaluating rounds.
\item [DRS:] support to detect, store and load looping solutions.
\end{description}

As these extensions are specific to each strategy, as we will see
next, they can be combined without major overheads.


\subsection{Tabling Operations}

We next introduce the pseudo-code for the main tabling operations
required to support the integration of the three strategies on top of
Yap.

We start with Fig.~\ref{fig_new_solution} showing the pseudo-code for
the \emph{new solution} operation. Initially, the operation simply
inserts the given solution \texttt{SOL} in the solution trie structure
for the given subgoal frame \texttt{SF} and, if the solution is new,
it updates the \texttt{SgFr\_new\_solutions} subgoal frame field to
\texttt{TRUE}. If DRS mode is enabled for the subgoal, it also marks
the newest solution found during the current round. We then implement
a local scheduling strategy and always fail at the end.

\begin{figure}[ht]
{\verbatimproperties
\begin{verbatim}
new_solution(solution SOL, subgoal frame SF) {
  if (solution_check_insert(SOL,SF) == TRUE) {               // new solution
    SgFr_new_solutions(SF) = TRUE
    if (DRS_mode(SF) && first_solution_in_current_round(SF) == NULL)
      first_solution_in_current_round(SF) = SOL
  }
  fail()                                                 // local scheduling
}
\end{verbatim}}
\caption{Pseudo-code for the \emph{new solution} operation}
\label{fig_new_solution}
\end{figure}

Figure~\ref{fig_tabled_call} shows the pseudo-code for the
\emph{tabled call} operation. New calls to tabled subgoals are
inserted into the table space by allocating the necessary data
structures. This includes allocating and initializing a new subgoal
frame to represent the given subgoal call (this is the case where the
state of \texttt{SF} is \texttt{ready}). In such case, the tabled call
operation then updates the state of \texttt{SF} to
\texttt{evaluating}; stores a new generator node\footnote{Generator,
  consumer and follower nodes are implemented as WAM choice points
  extended with some extra fields.}; and proceeds by executing the
current alternative.

\begin{figure}[ht]
{\verbatimproperties
\begin{verbatim}
tabled_call(subgoal call SC) {
  SF = subgoal_check_insert(SC)            // SF is the subgoal frame for SC
  if (SgFr_state(SF) == ready) {                              // first round
    SgFr_state(SF) = evaluating
    store_generator_node()
    goto execute(current_alternative())
  } else if (SgFr_state(SF) == loop_ready) {          // re-evaluation round
    SgFr_state(SF) = loop_evaluating
    store_generator_node()
    if (DRA_mode(SF))
      goto execute(first_looping_alternative())
    else
      goto execute(first_alternative())
  } else if (SgFr_state(SF) == evaluating or                  // first round
             SgFr_state(SF) == loop_evaluating) {     // re-evaluation round
    mark_current_branch_as_a_non_leader_branch(SF)
    if (DRA_mode(SF) or DRS_mode(SF))
      mark_current_branch_as_a_looping_branch(SF)
    if (DRE_mode(SF) && has_unexploited_alternatives(SF)) {
      store_follower_node()
      if (DRA_mode(SF) and SgFr_state(SF) == loop_evaluating)
        goto execute(next_looping_alternative())
      else
        goto execute(next_alternative())
    } else {
      store_consumer_node()
      goto consume_solutions(SF)
    }
  } else if (SgFr_state(SF) == complete)                // already evaluated
    goto completed_table_optimization(SF)
}
\end{verbatim}}
\caption{Pseudo-code for the \emph{tabled call} operation}
\label{fig_tabled_call}
\end{figure}

On the other hand, if the subgoal call is a repeated call, then the
subgoal frame is already in the table space, and three different
situations may occur. First, if the call is already evaluated (this is
the case where the state of \texttt{SF} is \texttt{complete}), the
operation consumes the available solutions by implementing the
\emph{completed table optimization} which executes compiled code
directly from the solution trie structure associated with the
completed call~\cite{RamakrishnanIV-99}.

Second, if the call is a first call in a re-evaluating round (this is
the case where the state of \texttt{SF} is \texttt{loop\_ready}), the
operation updates the state of \texttt{SF} to
\texttt{loop\_evaluating}; stores a new generator node; and proceeds
by re-executing the first looping alternative or the first matching
alternative, according to whether the DRA mode is enabled or disabled
for the subgoal.

Third, if the call is a consumer call (this is the case where the
state of \texttt{SF} is \texttt{evaluating} or
\texttt{loop\_evaluating}), the operation first marks the current
branch as a non-leader branch and, if in DRA or DRS mode, it also
marks the current branch as a looping branch. Next, if DRE mode is
enabled and there are unexploited alternatives (i.e., there is a
backtracking clause for the former call), it stores a follower node
and proceeds by re-executing the next looping alternative or the next
matching alternative, according to whether the DRA mode is enabled or
disabled for the subgoal. Otherwise, it stores a new consumer node and
starts consuming the available solutions.

To mark the current branch as a non-leader branch, we follow all
intermediate generator calls in evaluation up to the generator call
for frame \texttt{SF} and we mark them as non-leader calls (note that
the call at hand defines a new dependency for the current SCC). To
mark the current branch as a looping branch, we follow all
intermediate generator calls in evaluation up to the generator call
for frame \texttt{SF} and we mark the alternatives being evaluated or
the solutions being consumed by each call, respectively, as looping
alternatives or looping solutions.

Finally, we discuss in more detail how completion is detected.
Remember that after exploring the last matching clause for a tabled
call, we execute the \emph{fix-point check}
operation. Figure~\ref{fig_fixpoint_check} shows the pseudo-code for
its implementation.

\begin{figure}[ht]
{\verbatimproperties
\begin{verbatim}
fix-point_check(subgoal frame SF) {
  if (SgFr_is_leader(SF)){
    if (SgFr_new_solutions(SF)) {                       // start a new round
      SgFr_new_solutions(SF) = FALSE
      for each (subgoal in current SCC)
        SgFr_state(subgoal) = loop_ready
      SgFr_state(SF) = loop_evaluating
      if (DRA_mode(SF))
        goto execute(first_looping_alternative())
      else
        goto execute(first_alternative())
    } else {                             // complete subgoals in current SCC
      for each (subgoal in current SCC)
        SgFr_state(subgoal) = complete   
      goto completed_table_optimization(SF)              // local scheduling
  } else {                                              // not a leader call
    if (SgFr_new_solutions(SF))                   // propagate new solutions
      SgFr_new_solutions(current_leader(SF)) = TRUE  
    SgFr_new_solutions(SF) = FALSE                    // reset new solutions
    // local scheduling
    if (DRS_mode(SF))
      goto consume_looping_solutions_and_solutions_in_current_round(SF)
    else
      goto consume_solutions(SF)
  }
}
\end{verbatim}}
\caption{Pseudo-code for the \emph{fix-point check} operation}
\label{fig_fixpoint_check}
\end{figure}

The fix-point check operation starts by checking if the subgoal at
hand is a leader call. If it is leader and has found new solutions
during the current round, then the current SCC is scheduled for a
re-evaluation. If it is leader but no new solutions were found during
the current round, then we have reached a fix-point and thus, the
subgoals in the current SCC are marked as completed and the evaluation
proceeds with the completed table optimization. Otherwise, if the
subgoal is not a leader call, then it propagates the new solutions
information to the current leader of the SCC and starts consuming the
available solutions. If DRS mode is enabled, it only consumes the
looping solutions and the solutions found during the current round,
otherwise it consumes all solutions.


\section{Experimental Results}

To the best of our knowledge, Yap is now the first tabling engine that
integrates and supports the combination of different linear tabling
strategies. We have thus the conditions to better understand the
advantages and weaknesses of each strategy when used solely or
combined with the others. In what follows, we present initial
experiments comparing linear tabled evaluation with and without
support for the DRE, DRA and DRS strategies. The environment for our
experiments was a PC with a 2.83 GHz Intel(R) Core(TM)2 Quad CPU and 4
GBytes of memory running the Linux kernel 2.6.32-27-generic-pae with
Yap 6.0.7.

To put the performance results in perspective, we used two right
recursive definitions of the well-known $path/2$ predicate, that
computes the transitive closure in a graph, combined with several
different configurations of $edge/2$ facts. One path definition has
the recursive clause first and the other has the recursive clause
last.

{\verbatimproperties
\begin{verbatim}
path_first(X,Z) :- sld1, edge(X,Y), path_first(Y,Z), sld2.
path_first(X,Z) :- sld3, edge(X,Z), sld4.

path_last(X,Z) :- sld3, edge(X,Z), sld4.
path_last(X,Z) :- sld1, edge(X,Y), path_last(Y,Z), sld2.
\end{verbatim}}

\begin{wrapfigure}{!r}{6cm}
\centering
\includegraphics[width=6cm]{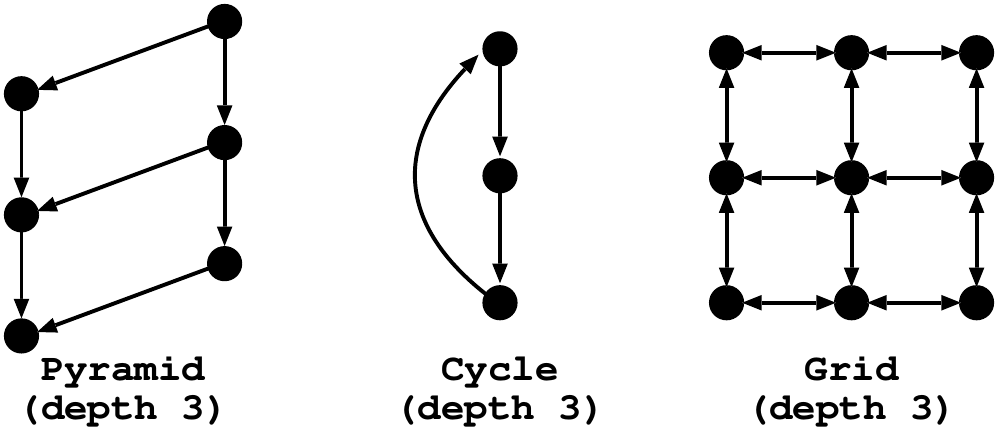}
\caption{An example of edge configurations}
\label{fig_edge_configurations}
\end{wrapfigure}

Regarding the edge facts, we used three configurations: a pyramid, a
cycle and a grid configuration (Fig.~\ref{fig_edge_configurations}
shows an example for each configuration). We experimented the pyramid
and cycle configurations with depths 1000, 2000 and 3000 and the grid
configuration with depths 20, 30 and 40. All experiments find all the
solutions for the problem. We chose these experiments because the
$path/2$ predicate implements a relatively easy to understand pattern
of computation and its right recursive definition creates several
inter-dependencies between tabled subgoals. Notice also that in the
definitions above we included four extra SLD (non-tabled) predicates
(the \texttt{sld1/0}, \texttt{sld2/0}, \texttt{sld3/0} and
\texttt{sld4/0} predicates) in order to measure how the mixing with
SLD computations can affect the base performance.

First, in Table~\ref{tab_times}, we show the execution time, in
milliseconds, for standard linear tabled evaluation with local
scheduling and the ratios comparing standard linear tabling against
DRE, DRA and DRS solely and combined strategies (\emph{All} means
DRE+DRA+DRS) for the two definitions of the $path/2$ predicate without
including the four extra SLD computations. Ratios higher than 1.00
mean that the respective strategies have a positive impact on the
execution time. The results obtained are the average of ten runs for
each configuration.

\begin{table}[ht]
\centering
\caption{Execution time, in milliseconds, for standard linear tabling
  with local scheduling and the respective ratios against the several
  strategies using the right recursive definition of the path problem
  (ratios in bold mean that the use of the respective strategies is
  better than not using some or all of them)}
\begin{tabular}{lrrrrrrrrr}
\hline\hline
\multicolumn{1}{c}{\multirow{2}{*}{\bf Strategy}} &
\multicolumn{3}{c}{\bf Pyramid} &
\multicolumn{3}{c}{\bf Cycle} &
\multicolumn{3}{c}{\bf Grid} \\ 
\cline{2-4}\cline{5-7}\cline{8-10} &
\multicolumn{1}{c}{\bf 1000} &
\multicolumn{1}{c}{\bf 2000} &
\multicolumn{1}{c}{\bf 3000} &
\multicolumn{1}{c}{\bf 1000} &
\multicolumn{1}{c}{\bf 2000} &
\multicolumn{1}{c}{\bf 3000} &
\multicolumn{1}{c}{\bf 20} &
\multicolumn{1}{c}{\bf 30} &
\multicolumn{1}{c}{\bf 40} \\
\hline
\multicolumn{10}{l}{\bf Recursive Clause First} \\
\bf{Standard} &       664 &     2,669 &     6,040
              &       377 &     1,522 &     3,400
              &       386 &     2,714 &    10,689 \\ \cline{2-10}
\bf{DRE}      & \bf{1.02} & \bf{1.01} & \bf{1.02}
              &     1.00  & \bf{1.01} & \bf{1.01}
              & \bf{1.02} &     1.00  &     1.00  \\
\bf{DRA}      & \bf{1.55} & \bf{1.51} & \bf{1.51}
              & \bf{1.22} & \bf{1.23} & \bf{1.21}
              & \bf{1.14} & \bf{1.09} & \bf{1.10} \\                       
\bf{DRS}      & \bf{1.01} &     1.00  & \bf{1.01}
              & \bf{1.21} & \bf{1.23} & \bf{1.22}
              & \bf{1.23} & \bf{1.27} & \bf{1.31} \\
\bf{DRE+DRA}  &     1.52  &     1.51  &     1.50 
              & \bf{1.24} &     1.23  &     1.20 
              & \bf{1.15} & \bf{1.10} &     1.06  \\
\bf{DRE+DRS}  &     1.01  &     1.01  &     1.00 
              & \bf{1.22} &     1.23  & \bf{1.22}
              &     1.22  &     1.23  &     1.23  \\
\bf{DRA+DRS}  &     1.54  & \bf{1.52} &     1.51 
              & \bf{1.56} & \bf{1.57} & \bf{1.52}
              & \bf{1.42} & \bf{1.42} & \bf{1.43} \\
\bf{All}      & \bf{1.56} & \bf{1.53} &     1.50 
              &     1.55  &     1.57  &     1.52 
              &     1.38  &     1.39  &     1.37  \\                  
\hline
\multicolumn{10}{l}{\bf Recursive Clause Last}  \\
\bf{Standard} &       673 &     2,775 &     6,216
              &       382 &     1,542 &     3,487 
              &       365 &     2,602 &    10,403 \\ \cline{2-10}
\bf{DRE}      &     0.99  & \bf{1.01} & \bf{1.01}
              & \bf{1.01} & \bf{1.01} & \bf{1.01}
              & \bf{1.02} & \bf{1.03} & \bf{1.03} \\
\bf{DRA}      & \bf{1.47} & \bf{1.49} & \bf{1.47}
              & \bf{1.24} & \bf{1.22} & \bf{1.22}
              & \bf{1.15} & \bf{1.13} & \bf{1.11} \\                       
\bf{DRS}      &     0.99  &     0.99  & \bf{1.01}
              & \bf{1.20} & \bf{1.21} & \bf{1.23}
              & \bf{1.21} & \bf{1.27} & \bf{1.30} \\
\bf{DRE+DRA}  & \bf{1.49} &     1.34  &     1.43
              &     1.24  &     1.22  &     1.22
              &     1.14  &     1.12  &     1.10  \\
\bf{DRE+DRS}  &     1.00  &     0.99  &     1.01
              & \bf{1.23} &     1.22  &     1.23
              & \bf{1.22} &     1.27  &     1.30  \\ 
\bf{DRA+DRS}  &     1.47  &     1.47  &     1.46
              & \bf{1.55} & \bf{1.54} & \bf{1.53}
              & \bf{1.42} & \bf{1.43} & \bf{1.43} \\
\bf{All}      &     1.49  &     1.48  &     1.09
              &     1.48  & \bf{1.56} & \bf{1.55}
              &     1.42  & \bf{1.44} & \bf{1.45} \\
\hline\hline
\end{tabular}
\label{tab_times}
\end{table}

In addition to the results presented in Table~\ref{tab_times}, we also
collected several statistics regarding important aspects of the
evaluation (not fully presented here due to lack of space). In
Table~\ref{tab_stats}, we show some of these statistics for standard
linear tabled evaluation and the ratios against the several strategies
for the particular evaluation of the grid configuration with depth
40. The \emph{Alts} column shows the number of alternatives explored
during the evaluation, the \emph{Sols} column shows the number of
solutions consumed by generator nodes corresponding to non-leader
subgoals, and the \emph{SLD} columns show the number of times each
extra SLD predicate is executed.

\begin{table}[ht]
\centering
\caption{Statistics for standard linear tabling and the respective
  ratios against the several strategies for the grid configuration
  with depth 40 (ratios in bold mean that the use of the respective
  strategies is better than not using some or all of them)}
\begin{tabular}{lrrrrrr}
\hline\hline
\multicolumn{1}{c}{\multirow{2}{*}{\bf Strategy}} &
\multicolumn{1}{c}{\multirow{2}{*}{\bf Alts}} & 
\multicolumn{1}{c}{\multirow{2}{*}{\bf Sols}} & 
\multicolumn{4}{c}{\bf SLD Computations} \\
\cline{4-7} & & &
\multicolumn{1}{c}{\texttt{\bf sld1/0}} &
\multicolumn{1}{c}{\texttt{\bf sld2/0}} &
\multicolumn{1}{c}{\texttt{\bf sld3/0}} &
\multicolumn{1}{c}{\texttt{\bf sld4/0}} \\
\hline
\multicolumn{3}{l}{\bf Recursive Clause First} \\
\bf{Standard} &   70,403  & 50,015,215  &   35,202  & 200,974,309  &    35,201  &   149,757  \\ \cline{2-7}
\bf{DRE}      & \bf{1.05} &   \bf{1.04} & \bf{1.05} &    \bf{1.04} &  \bf{1.05} &  \bf{1.04} \\
\bf{DRA}      & \bf{1.91} &       1.00  &     1.00  &    \bf{1.05} & \bf{21.99} & \bf{12.00} \\
\bf{DRS}      &     1.00  &  \bf{19.55} &     1.00  &    \bf{1.29} &      1.00  &      1.00  \\
\bf{DRE+DRA}  &     1.06  &       1.04  &     1.05  &    \bf{1.10} &      1.07  &      1.11  \\
\bf{DRE+DRS}  &     1.05  &      19.55  &     1.05  &    \bf{1.33} &      1.05  &      1.04  \\
\bf{DRA+DRS}  &     1.91  &      19.55  &     1.00  &    \bf{1.38} &     21.99  &     12.00  \\
\bf{All}      &     1.06  &      19.55  &     1.05  &    \bf{1.43} &      1.07  &      1.11  \\
\hline
\multicolumn{3}{l}{\bf Recursive Clause Last} \\
\bf{Standard} &   67,204  & 48,080,300  & 48,602 & 352,277,129  &    48,602  &   205,920  \\ \cline{2-7}
\bf{DRE}      &     1.00  &       1.00  &   1.00 &        1.00  &      1.00  &      1.00  \\
\bf{DRA}      & \bf{1.91} &       1.00  &   1.00 &    \bf{1.05} & \bf{20.99} & \bf{11.50} \\
\bf{DRS}      &     1.00  &  \bf{18.79} &   1.00 &    \bf{1.29} &      1.00  &      1.00  \\
\bf{DRE+DRA}  &     1.91  &       1.00  &   1.00 &        1.05  &     20.99  &     11.50  \\
\bf{DRE+DRS}  &     1.00  &      18.79  &   1.00 &        1.29  &      1.00  &      1.00  \\
\bf{DRA+DRS}  &     1.91  &      18.79  &   1.00 &    \bf{1.38} &     20.99  &     11.50  \\
\bf{All}      &     1.91  &      18.79  &   1.00 &        1.38  &     20.99  &     11.50  \\
\hline\hline
\end{tabular}
\label{tab_stats}
\end{table}

Analyzing the general picture of Table~\ref{tab_times}, the results
show that, for most of these experiments, DRE evaluation has no
significant impact in the execution time. On the other hand, the
results indicate that the DRA and DRS strategies are able to
effectively reduce the execution time for most of the experiments,
when compared with standard evaluation, and that by combining both
strategies it is possible to obtain even better results. We next
discuss in more detail each strategy.

\begin{description}
\item[DRE:] for most of these configurations, DRE has no significant
  impact. For the configurations with the recursive clause last and
  the configurations without loops (i.e., without inter-dependencies
  between subgoals), like the pyramid configurations, it is not
  applicable and thus no followers nodes are ever stored. For the
  cycle configurations the number of followers is also very reduced,
  maximum 3 followers, and thus its impact is insignificant. For the
  grid configurations with the recursive clause first, the results
  obtained are the most interesting. For example, in
  Table~\ref{tab_stats} for the recursive clause first, DRE executes
  less alternatives (ratio 1.05) and consumes less solutions on
  non-leader generator nodes (ratio 1.04) than standard evaluation,
  but even so the impact on the execution time is minimal.
\item[DRA:] the results for DRA evaluation show that the strategy of
  avoiding the exploration of non-looping alternatives in
  re-evaluation rounds is quite effective in general. The results also
  show that DRA is more effective for programs without loops (thus
  without looping alternatives), like the pyramid configurations, than
  for programs with larger SCCs, like the cycle and grid
  configurations. For the pyramid and grid configurations, the total
  number of alternatives explored by the other strategies is around 2
  times the total number of alternatives explored with DRA. For the
  cycle configurations, this number is around 1.5 times the number
  with DRA evaluation. For example, in Table~\ref{tab_stats}, we can
  observe that standard evaluation explores 1.91 times more
  alternatives (respectively 33,601 and 32,002 more alternatives for
  the recursive clause first and last) than DRA evaluation for the
  grid configuration with depth 40.
\item[DRS:] the results for DRS evaluation show that the strategy of
  avoiding the consumption of non-looping solutions in re-evaluation
  rounds is quite effective for programs that can benefit from it,
  like the cycle and grid configurations, and do not introduces
  relevant costs for programs that cannot benefit from it, like the
  pyramid configurations. Notice that the pyramid configurations only
  execute one re-evaluation round per SCC and that we only take
  advantage of DRS evaluation starting from the second re-evaluation
  round. For the cycle and grid configuration, DRS optimization is
  used several times because these configurations create a single huge
  SCC including all subgoal calls. For example, in
  Table~\ref{tab_stats}, DRS consumes 47,456,815 (ratio 19.55) and
  45,521,900 (ratio 18.79) less solutions than standard evaluation for
  the recursive clause first and last, respectively.
\end{description}

Regarding the combination of the strategies, in general, our
statistics show that the best of each world is always present in the
combination. By analyzing the results in Table~\ref{tab_times}, we can
conclude that, for these experiments, by combining the DRA and DRS
strategies it is possible to reduce even further the execution time of
the evaluation, and in most cases this reduction is higher than the
sum of the reductions obtained with each strategy individually. In
particular, Table~\ref{tab_stats} shows the same number of solutions
and alternatives for DRA+DRS that the respective DRS and DRA
strategies obtain when used solely. This clearly shows the potential
of our framework and suggests that the overhead associated with this
combination is negligible. When DRE is present, the results are, in
general, worse than the results obtained with the DRA/DRS strategies
solely. In Table~\ref{tab_stats} we can observe that, for the DRE+DRA
combination, the number of the alternatives explored is far more
higher than the DRA used solely and that, for the DRE+DRS combination,
the non consumed solutions for DRE used solely are included on the non
consumed solutions of the DRS optimization. So, for this particular
configurations, DRE is not fully orthogonal to DRA and DRS.

Table~\ref{tab_stats} also shows the number of times each extra SLD
predicate is executed for the grid configuration with depth 40. We can
read these numbers as an estimation of the performance ratios that we
will obtain if the execution time of the corresponding SLD predicate
clearly overweights the execution time of the other tabled and
non-tabled computations. In particular, the \texttt{sld2/0} predicate
(placed at the end of the recursive clause) is the one that can
potentially have a greater influence in the performance ratios as it
clearly exceeds all the others in the number of executions.

In general, these ratios show that by mixing tabled with non-tabled
computations, our framework can achieve similar and, for some cases,
even better results than the ones presented in
Table~\ref{tab_times}. In particular, the ratios for the
\texttt{sld2/0} predicate (the one with greater influence) are very
similar to the ratios in Table~\ref{tab_times} and for DRA evaluation,
the ratios for the \texttt{sld3/0} and \texttt{sld4/0} predicates are
excellent (around 22 and 12, respectively), showing that intertwining
SLD computations with linear tabling can affect positively the base
performance.


\section{Conclusions}

We presented a new strategy for linear tabled evaluation of logic
programs, named DRS, and a framework that integrates and supports the
combination of DRS with two other (DRE and DRA) linear tabling
strategies. We discussed how these strategies can optimize different
aspects of a tabled evaluation and we presented the relevant
implementation details for their integration on top of the Yap system.

Our experiments for DRS evaluation showed that the strategy of
avoiding the consumption of non-looping solutions in re-evaluation
rounds can be quite effective for programs that can benefit from it,
with insignificant costs for the other programs. Preliminary results
for the combined framework were also very promising. In particular,
the combination of DRA with DRS showed the potential of our framework
to reduce even further the execution time of a linear tabled
evaluation.

Further work will include adding new strategies/optimizations to our
framework, such as the ones proposed in~\cite{Zhou-08}, and exploring
the impact of applying our strategies to more complex problems,
seeking real-world experimental results, allowing us to improve and
consolidate our current implementation.


\section*{Acknowledgments}

This work has been partially supported by the FCT research projects
HORUS (PTDC/EIA-EIA/100897/2008) and LEAP (PTDC/EIA-CCO/112158/2009).


\bibliographystyle{acmtrans}
\bibliography{references}


\end{document}